\documentclass[aps, prd, twocolumn, amsmath, floats,floatfix, superscriptaddress, nofootinbib, showkeys]{revtex4}

\usepackage{amssymb}
\usepackage{amsmath}
\usepackage{verbatim}
\usepackage{mathrsfs}
\usepackage{amsfonts}
\usepackage{latexsym}
\usepackage{epsfig}
\usepackage{color}
\usepackage{graphicx,subfigure}
\usepackage{units}
\usepackage{envmath}
\usepackage{natbib}
\usepackage{ctable}
\usepackage{soul}
\usepackage{ulem}

\begin{document}


\definecolor{orange}{rgb}{0.9,0.45,0}

\newcommand{\re}{\mbox{Re}}
\newcommand{\im}{\mbox{Im}}
\newcommand{\mg}[1]{\textcolor{red}{MG: #1}} 
\newcommand{\saeed}[1]{\textcolor{blue}{SF: #1}}   
\newcommand{\ab}[1]{\textcolor{orange}{AB: #1}}   
\newcommand{\jtf}[1]{\textcolor{violet}{JTF: #1}}   

\def\CovDev{D}
\def\Res{{\mathcal R}}
\def\Gammaflat{\hat \Gamma}
\def\metricflat{\hat \gamma}
\def\Dflat{\hat {\mathcal D}}
\def\part_n{\partial_\perp}

\def\Lie{\mathcal{L}}
\def\A{\mathcal{X}}
\def\Aphi{\A_{\phi}}
\def\hAphi{\hat{\A}_{\phi}}
\def\E{\mathcal{E}}
\def\Ham{\mathcal{H}}
\def\M{\mathcal{M}}
\def\R{\mathcal{R}}
\def\p{\partial}

\def\hg{\hat{\gamma}}
\def\hA{\hat{A}}
\def\hD{\hat{D}}
\def\hE{\hat{E}}
\def\hR{\hat{R}}
\def\hcA{\hat{\mathcal{A}}}
\def\hDelt{\hat{\triangle}}

\def\na{\nabla}
\def\dif{{\rm{d}}}
\def\non{\nonumber}
\newcommand{\erf}{\textrm{erf}}

\renewcommand{\t}{\times}

\long\def\symbolfootnote[#1]#2{\begingroup%
\def\thefootnote{\fnsymbol{footnote}}\footnote[#1]{#2}\endgroup}


\title{The Integrated Sachs-Wolfe Effect in 4D Einstein-Gauss-Bonnet Gravity}
\author{Mina Ghodsi Y.}
\email{m.ghodsi.y@gmail.com}
\affiliation{Department of Physics, Shahid Beheshti University, G.C., Evin, Tehran 19839, Iran}

\author{Aryan Behnamfard}
\email{aryan.bnf@gmail.com}
\affiliation{Department of Physics, Shahid Beheshti University, G.C., Evin, Tehran 19839, Iran}

\author{Saeed Fakhry} 
\email{s_fakhry@sbu.ac.ir}
\affiliation{Department of Physics, Shahid Beheshti University, G.C., Evin, Tehran 19839, Iran}

\author{Javad T. Firouzjaee} 
\email{firouzjaee@kntu.ac.ir}
\affiliation{Department of Physics, K.N. Toosi University of Technology, P.O. Box 15875-4416, Tehran, Iran}
\affiliation{School of Physics, Institute for Research in Fundamental Sciences (IPM), P.O. Box 19395-5531, Tehran, Iran}

\date{\today}


\begin{abstract} 

A novel 4-dimensional Einstein-Gauss-Bonnet (4D EGB) gravity has been proposed that asserts to bypass the Lovelock's theorem and to result in a non-trivial contribution to the gravitational dynamics in four-dimensional spacetime. In this work, we study the integrated Sachs-Wolfe (ISW) effect in the 4D EGB model. For this purpose, we calculate the evolution of the gravitational potential and the linear growth factor as a function of redshift for the 4D EGB model and compare it with the corresponding result obtained from the $\Lambda$-cold dark matter ($\Lambda$CDM) model. We also calculate the ISW-auto power spectrum and the ISW-cross power spectrum as functions of cosmic microwave background (CMB) multipoles for the 4D EGB model and compare those with the one obtained from the $\Lambda$CDM model. To do this, we use the strongest constraint on the coupling parameter proposed for the 4D EGB model. Additionally, to calculate the ISW effect for the 4D EGB model, we employ three large-scale structure surveys from different wavelengths. The results exhibit that the ISW effect in the 4D EGB model is higher than the one obtained from the $\Lambda$CDM model. Hence, we show that the 4D EGB model can amplify the amplitude of the ISW power spectrum, which can be considered as a relative advantage of the 4D EGB model comparing the $\Lambda$CDM one. Also, we indicate that the deviation from the $\Lambda$CDM model is directly proportional to the value of the dimensionless coupling parameter $\beta$.

\end{abstract}


\maketitle

\vspace{0.8cm}

\section{Introduction}\label{sec:intro}
Measurements of cosmic microwave background (CMB) anisotropies have disclosed that the cosmos can be described remarkably by a six-parameter model, with the current energy content dominated by cold dark matter (CDM) and dark energy consistent with a cosmological constant as the standard model of cosmology \cite{wmap, plank}. CMB photons along their path to the earth are subject to various interactions, namely weak gravitational lensing, integrated Sachs-Wolfe and Rees-Sciama effect, and Shapiro effect, which depend directly on the gravitational field.
If Photons cross a time-evolving gravitational potential, the energy gained by photons falling into a potential well cannot cancel out the energy loss as photons climb out of the well \cite{Sachs:1967er, 0801.0642}. For photons passing through a growing potential well, the light is redshifted, and it seems blueshifted if the potential well is decaying. This effect is known as the integrated Sachs-Wolfe (ISW) effect \cite{2007.04968}. This effect can give an important observable that constrains dark energy models. In other words, the ISW effect leads to a new direction in observational cosmology taken for constraining the cosmological models \cite{Cai:2013toa, Nishizawa:2014vga, Manzotti:2014kta, Mostaghel:2018pia, Beck:2018owr, Dong:2020fqt}.
It is known that the simplest way to study the so-called dark energy, responsible for the observed cosmic acceleration, is to postulate the existence of a cosmological constant. Nevertheless, this explanation has some serious problems; above all, the dark energy scale is incredibly smaller than the typical particle physics vacuum energy \cite{Weinberg:1988cp}. This is one of the reasons why a lot of alternative models have been proposed, providing different models of dark energy \cite{Peebles:2002gy, Padmanabhan:2002ji, Copeland:2006wr, Lombriser:2019jia}. These discussions (similar problems for dark matter) lead to a crucial idea of modifying the general relativity (GR), which cannot be well explained in the framework of GR plus $\Lambda$-cold dark matter ($\Lambda$CDM), where $\Lambda$ is the cosmological constant.
There is an important way to add dark energy to the GR, i.e., promoting the cosmological constant to a dynamical field. This gives that the basic theory governing the gravitational dynamics of the Universe may not be GR and could be an alternative gravitational scenario, which can help to realize the dark side better. One way to generalize the Ricci scalar in the Einstein-Hilbert (EH) action is to add some higher-order terms made from Ricci and Riemann tensors. Lovelock's terms will give a well-known model, which cancels the ghost degrees of freedom \cite{Lovelock1, Lovelock2, Lovelock3} if the following four assumptions can be satisfied: ({\it i}) metricity, ({\it ii}) diffeomorphism invariance, ({\it iii}) second-order equations of motion, ({\it iv}) $(3+1)$-dimensional spacetime. One of the non-trivial Lovelock terms is the Gauss-Bonnet (GB) invariant which is a modification of GR, which leads to the GB gravity \cite{Gauss-Bonnet1, Gauss-Bonnet2} and satisfies the former assumptions in higher dimensions, i.e., in $D>4$. Since GB invariant is a total derivative in 4D spacetime, it can not contribute to gravitational dynamics. Interest in the dimensional regularization of gravity with the GB term was motivated by the development of 4D EGB gravity \cite{Glavan:2019inb}. Since due to the presence of an overall factor $(D-4)$, the GB invariant vanishes in 4D, authors in Ref.~\cite{Glavan:2019inb} solved this issue by rescaling the coupling constant $ \alpha\rightarrow \alpha/(D-4) $ in order to produce non-trivial contributions to gravitational dynamics. Consequently, many aspects of this theory have been investigated in black holes and compact objects physics~\cite{Kumar:2020owy, Zhang:2020qew, Aragon:2020qdc, Yang:2020czk, Lin:2020kqe, Wei:2020poh, Konoplya:2020qqh, Konoplya:2020juj, Yang:2020jno, Heydari-Fard:2021ljh, HosseiniMansoori:2020yfj, Wei:2020ght}, and in cosmology~\cite{Li:2020tlo, Aoki:2020iwm, MohseniSadjadi:2020qnm, Narain:2020qhh, Aoki:2020ila, Shahidi:2021rnu}.
In this work, we calculate the ISW effect in the 4D EGB gravity. In this respect, the outline of the work is as follows. In Sec.~\ref{sec:ii}, we review the theoretical framework of the 4D EGB gravity. Then, in Sec.~\ref{sec:iii}, we discuss the cosmological consequences of the 4D EGB gravity concerning the homogeneous and inhomogeneous Universe. Moreover, in Sec.~\ref{sec:iv}, we calculate the ISW effect in the 4D EGB model, and compare it with the corresponding result obtained from the $\Lambda$CDM model. Finally, we scrutinize the results and summarize the findings in Sec.~\ref{sec:v}.
\section{Theoretical Framework}\label{sec:ii}
The Lovelock's theorem ensures that for a gravitational theory with four assumptions mentioned in Sec.~\ref{sec:intro}, the EH term together with the cosmological constant term lead to the unique theory for any generic form of symmetric rank-2 tensor $g_{\mu \nu}$. In other words, this theorem guarantees the uniqueness of Einstein's field equations in 4D spacetime. The well-known EH action with cosmological constant is
\begin{equation}
S_{\rm EH} = \int d^Dx \sqrt{-g}\,\kappa^{2}\,(R - 2\Lambda),
\end{equation}
where $D$ is the dimension of spacetime, $\kappa$ is Einstein's gravitational constant, $\Lambda$ is the cosmological constant, and $R$ is the scalar curvature. Also, $g={\rm det}(g_{\mu \nu})$ is the determinant of the metric tensor matrix. Nevertheless, in higher-dimensional spacetime, i.e., $D>4$, the EH action with second-order field equations is no longer unique. In this regard, a unique action, i.e., the GB action, has been proposed in higher dimensions that include quadratic corrections of the curvature tensors~\cite{Lovelock1, Lovelock2, Lovelock3, Zwiebach1985}. The GB action in $D$-dimensional spacetime is
\begin{equation}
S_{\rm GB} = \int d^Dx \sqrt{-g}\,\alpha\,\mathcal{G},
\end{equation}
where $\alpha$ is the GB coupling parameter. In this action, $\mathcal{G}$ is the GB invariant and has the following form
\begin{equation}
\mathcal{G} = R^{\mu \nu \rho \sigma} R_{\rm \mu \nu \rho \sigma} - 4R^{\mu \nu} R_{\rm \mu \nu} + R^{2},
\end{equation}
where $R_{\mu \nu \rho \sigma}$ is the Riemann curvature tensor, $R_{\mu \nu}$ is the Ricci curvature tensor and $R^{2}$ is the squared of the scalar curvature. In 4-dimensional spacetime, the GB invariant is a total derivative and does not alter the dynamics of the gravitational system~\cite{Felice}. In other words, variation of the GB action with respect to the metric yields
\begin{equation}
\frac{g_{\rm \mu \nu}}{\sqrt{-g}} \frac{\delta S_{\rm GB}}{\delta g_{\rm \mu \nu}} = \frac{(D-4)}{2}\alpha \mathcal{G},
\end{equation}
which is identically zero in $D=4$. While in higher dimensions, i.e., $D>4$, it does not vanish and appears as a dynamical term in the field equations. Recently, 4D EGB has been proposed to rescale GB coupling parameter as $\alpha \rightarrow \alpha / (D-4)$ in a way that it cancels the factor of $(D-4)$ in the GB action~\cite{Glavan:2019inb}. Under this assumption in the limit of $D \rightarrow 4$, one can expect to have a non-negligible dynamical contribution of the GB action in the field equations. Therefore, by combining the EH action, the rescaled GB one, and the action of matter field, $S_{\rm m}$, one can achieve the total action that can intelligently circumvent the Lovelock's theorem and its consequences. With these considerations, the total action can be specified as
\begin{multline} \label{tot_action}
S = \int d^Dx \sqrt{-g} \{ \kappa^2 (R - 2 \Lambda) + \frac{\alpha}{D-4} \mathcal{G} + \mathcal{L}_{\rm m} \},
\end{multline}
where $\mathcal{L}_{\rm m}$ is the Lagrangian of matter field. As can be seen, by rescaling the GB coupling parameter, the factor of $(D-4)$ in the GB term will be eliminated. Hence, variation of the total action with respect to the metric leads to the modified Einstein's field equations, namely
\begin{multline}\label{VarLE$GB$}
2 \kappa^2 (G_{\rm \mu \nu} + \Lambda g_{\rm \mu \nu}) + \frac{\alpha}{d-4} (2 R R_{\rm \mu \nu} - 4 R_{\rm \mu}^{\alpha} R_{\rm \nu \alpha} \\
- 4R^{\alpha \beta} R_{\rm \mu \alpha \nu \beta} - g_{\rm \mu \nu} \mathcal{G}) = T_{\rm \mu \nu},
\end{multline}
where $T_{\rm \mu \nu}$ is the energy-momentum tensor of the matter field and can be calculated by the following defination
\begin{equation}
T_{\rm \mu \nu} =-\frac{2}{\sqrt{-g}}\frac{\delta(\sqrt{-g}\mathcal{L}_{\rm m})}{\delta g^{\rm \mu \nu}}.
\end{equation}

The second parenthesis in Eq.~(\ref{VarLE$GB$}) is known as the GB tensor, which indicates the contribution of the GB action in the equations of motion while considering the rescaled GB coupling parameter. Specifically, the total action introduced in Eq.~(\ref{tot_action}) represents a classical modified theory of gravity that will reduce to the general relativity without any Ostrogradski instability if $\alpha\rightarrow 0$. Moreover, the number of degrees of freedom remains constant within this limit. Furthermore, by applying the Bianchi identity to the GB tensor, one can quickly investigate the conservation of energy-momentum tensor by the following relation
\begin{equation} \label{EM-full}
\nabla_{\mu} T^{\mu \nu} = 0,
\end{equation}
where $\nabla_{\mu}$ represents the covariant derivative. Note that the conservation of energy-momentum tensor is trivial in the absence of non-minimal coupling between the matter and geometry. In the following, we will discuss the cosmological implications of the 4D EGB model. 
\section{Cosmology}\label{sec:iii}
\subsection{Homogeneous Universe}
It is known that the Universe is assumed to be isotropic and homogeneous on sufficiently large scales. In this regard, we intend to study the cosmological consequences of the 4D EGB theory by applying the homogeneous and flat spacetime to the geometry of the Universe. For this purpose, one can adopt the Friedmann-Lemaitre-Robertson-Walker (FLRW) metric as
\begin{equation}\label{frw}
ds^2 = a^2(\tau) \eta_{\mu \nu} dx^{\mu} dx^{\nu},
\end{equation}
where $\eta_{\mu \nu}=\rm diag(-1, +1, +1,\cdots)$ is the Minkowski metric and $a(\tau)$ is the scale factor that is as a function of conformal time $\tau$. In D-dimensional spacetime, one can expand relation~(\ref{frw}) as
\begin{equation}\label{exp_frw}
ds^2=-a^{2}(\tau)d\tau^{2} + a^{2}(\tau) ( dx_1^{2} + dx_2^{2} + ... + dx_{D-1}^{2} ).
\end{equation}
Therefore, by inserting relation~(\ref{exp_frw}) into Eq.~(\ref{VarLE$GB$}) the Riemann curvature tensor, the Ricci curvature tensor, and the scalar curvature can be calculated to be
\begin{equation}
\begin{array}{l}
R_{i0i0} = -a \ddot{a}+ \dot{a}^2,	\\ \\
R_{ijij} = \dot{a}^2	, \\ \\
R_{00} = - (D-1)\dfrac{a \ddot{a} - \dot{a}^2}{a^2}, \\ \\
R_{ii} = \dfrac{(D-3) \dot{a}^2 + a \ddot{a}}{a^2}, \\ \\
R = \dfrac{(D-1) [2 a \ddot{a} + (D-4) \dot{a}^2]}{a^4},
\end{array}
\end{equation}
where dot indicates derivative with respect to the conformal time $\tau$. On the other hand, the perfect fluid is the best approximation proposed for the energy-momentum tensor in the homogeneous and isotropic Universe, which has the following form
\begin{equation}
T_{\mu \nu} = (\rho + p) u_{\mu} u_{\nu} - p g_{\mu \nu}.
\end{equation}
In this relation, $\rho$, $p$ and $u^{\mu}$ are the energy density, the pressure and the four-velocity vector field of the cosmological fluid, respectively. Additionally, we have assumed the barotropic equation of state (EoS) for the matter field, i.e. $p=\omega\rho$, where $\omega$ is a dimensionless parameter.
Under the above assumptions, the Friedmann and Raychaudhuri equations for the 4D EGB gravity can be specified as
\begin{equation}
6 \mathcal{H}^2 (\alpha \mathcal{H}^2 + a^2 \kappa^2) = a^4 (2 \kappa^2 \Lambda + \rho) ,
\label{Friedmann-eq}
\end{equation}
and
\begin{equation}
4 \mathcal{\dot{H}} (2 \alpha \mathcal{H}^2 + a^2 \kappa^2) - 2 \mathcal{H}^2 (\alpha \mathcal{H}^2 - a^2 \kappa^2) = a^4 (2 \kappa^2 \Lambda - p),
\end{equation}
where $\mathcal{H} = \dot{a}/a$ is the Hubble parameter. In the Minkowski spacetime, one can easily show that the energy-momentum tensor is conserved, namely
\begin{equation} \label{EMC}
\nabla_{\mu} T^{\mu}_{\nu} = \partial_{\mu} T^{\mu}_{\nu} + \Gamma^{\mu}_{\mu \lambda} T^{\lambda}_{\nu} - \Gamma^{\lambda}_{\mu \nu} T^{\mu}_{\lambda} = 0.
\end{equation}
Hence, by taking $\nu = 0$ in Eq.~(\ref{EMC}) the continuity equation can be obtained as follow
\begin{equation}
\dot{\rho} + 3 \mathcal{H} (\rho + p) = 0.
\end{equation}
On the other hand, the dimensionless parameter of the EoS for the ordinary non-relativistic matter with zero pressure is $\omega_{\rm m} = 0$, whereas it takes the form of $\omega_{\rm r} = 1/3$ for the case of radiation. Therefore, for a Universe that includes non-relativistic matter and radiation, density and pressure can be defined as follows
\begin{equation}
\rho = \rho_{\rm m} + \rho_{\rm r}=6 \kappa^2 H_0^2 (\dfrac{\Omega_{\rm r0}}{a^4} + \dfrac{\Omega_{\rm m0}}{a^3}), \hspace*{0.5cm} p = p_{\rm r} = \frac{\rho_{\rm r}}{3},
\label{rho-p}
\end{equation}
where $H_0$ is the present-time Hubble parameter, and $\rho_{\rm m}$ and $\rho_{\rm r}$ denote the matter and radiation contribution of the energy density, respectively. Also, $\Omega_{\rm r0}$ is the density parameter for radiation, and $ \Omega_{\rm m0}$ the density parameter for matter, and $ \Omega_{\Lambda 0}$ is the density parameter for dark energy at the present-time Universe. By inserting Eq.~(\ref{rho-p}) together with the following dimesionless parameters
\begin{equation}
\begin{array}{l}
E(a)=\dfrac{\mathcal{H}(a)}{H_0},	\kern 2pc	\beta = \dfrac{H_0^2 \alpha}{\kappa^2}, \kern 2pc \Omega_{\Lambda 0} = \dfrac{\Lambda}{3 H_0^2},
\label{parameters}
\end{array}
\end{equation}
one can rewrite Eq.~(\ref{Friedmann-eq}) to be
\begin{equation}
E^2(a) + \dfrac{\beta}{a^2} E^4(a)= a^2 (\dfrac{\Omega_{\rm r0}}{a^4} + \dfrac{\Omega_{\rm m0}}{a^3} + \Omega_{\Lambda 0}).
\label{friedmann-E-eq}
\end{equation}
This relation explains the evolution of the homogeneous Universe while considering the 4D EGB model. Moreover, at the present-time Universe (i.e., at $a=1$), a relation between those density parameters can be extracted from Eq.~(\ref{friedmann-E-eq}), namely
\begin{equation}
\Omega_{\rm r0} + \Omega_{\rm m0} + \Omega_{\rm \Lambda 0} - \beta = 1 .
\end{equation}
We have strong evidence that the Universe is Euclidean, and total density parameter is $\Omega \equiv \Omega_{\rm r}+\Omega_{\rm m}+\Omega_{\Lambda} = 1$ \cite{Dodelson2020}. Correspondingly, the value of the dimensionless coupling parameter, $\beta$, should be very small. Various approaches have been performed to find the observational constraint on $\beta$. These constraints come from gravitational waves, cosmic microwave background, baryon acoustic oscillation, type Ia supernovae, and orbit of Mercury~\cite{2006.15017, 2006.16751, 2103.12358}. By analyzing the data obtained from these phenomena, the strongest constraint on the 4D EGB coupling parameter has been obtained as $\alpha = (2.69 \pm 11.67) \times 10^{48}~ \rm eV^{-2}$, which corresponds to $\beta = (1.2 \pm 5.2) \times 10^{-17}$ \cite{2103.12358}. In this work, we have considered these strong constraints placed on the coupling parameter to calculate the ISW effect in 4D EGB model.

Up to here, we have discussed the evolution of the homogeneous Universe within the context of the 4D EGB model, while a clearer picture of how structures could have been formed and evolved is achieved by considering a Universe that is inhomogeneous. In the next section, we will study the evolution of the inhomogeneous Universe for the 4D EGB model within the context of the perturbation theory and will discuss its cosmological consequences.

\subsection{Inhomogeneous Universe}
There are different kinds of structures in the Universe like galaxies, galaxy clusters, and filaments. Although the Universe seems homogeneous on large scales, one may need some inhomogeneities in small scales to explain the formation conditions of these structures. It should be noted that these inhomogeneities in the cosmic fluid have to be negligible. For this purpose, we have employed a perturbed metric as
\begin{equation}
g_{\rm \mu \nu} = \bar{g}_{\rm \mu \nu} + \delta g_{\rm \mu \nu}, \hspace*{0.5cm} |\delta g_{\rm \mu \nu}|\ll1,
\end{equation}
where $\bar{g}_{\mu \nu}$ is the background metric and $\delta g_{\mu \nu}$ is the spacetime perturbations. Furthermore, in order to maintain the general covariance, a convenient gauge must be considered. To achieve this task, we have chosen the Newtonian gauge that leads to the following line element of spacetime
\begin{equation}
ds^2 = a^2(\tau) [-(1+2\phi) d\tau^2 + (1-2\psi) d \mathbf{x} ^2],
\end{equation}
which is known as perturbed conformal FLRW metric. In this relation, $\phi(\tau, \mathbf{x})$ and $\psi(\tau, \mathbf{x})$ are special combinations of metric perturbations and do not change under a coordinate transformation. Therefore, one can easily calculate the corresponding perturbed connection coefficients as
\begin{equation}
\begin{array}{l}
\Gamma^0_{00} = \mathcal{H} + \dot{\phi},	\kern 2.3pc	\Gamma^0_{0i} = \partial_i \phi	, \\	\\
\Gamma^i_{00} = \partial_j \phi \delta^{ij},	\kern 2pc	\Gamma^i_{j0} = (\mathcal{H} - \dot{\psi}) \delta^i_j ,	\\	\\
\Gamma^0_{ij} = \{-2\mathcal{H} (\phi + \psi) + \mathcal{H} - \dot{\psi}\}\delta_{ij}, \\ \\
\Gamma^i_{jk} = -2 \delta^i_{(j} \partial^{\ }_{k)} \psi + \delta_{jk} \delta^{il} \partial_l \psi ,
\label{Pert-Chr}
\end{array}
\end{equation}
where $\delta_{ij}$ denotes the Kronecker delta. Additionally, one can consider the small perturbations of the stress-energy tensor as
\begin{equation}\label{energy-momentum-full}
T^{\mu}_{\nu} = \bar{T}^{\mu}_{\nu} + \delta T^{\mu}_{\nu},
\end{equation}
where $\bar{T}_{\mu \nu}$ represents the energy-momentum tensor in a homogeneous and isotropic Universe and takes the form of a perfect fluid. Also, $\delta T^{\mu}_{\nu}$ is the perturbation of the energy-momentum tensor that is
\begin{equation}
\delta T^{\mu}_{\nu} = (\delta \rho + \delta p) \bar{u}^{\mu} \bar{u}_{\nu} + (\bar{\rho} + \bar{p})(\delta u^{\mu} \bar{u}_{\nu} + \bar{u}^{\mu} \delta u_{\nu}) - \delta p \delta^{\mu}_{\nu}
\end{equation}
where $\delta \rho$, $\delta p$ and $\delta u^{\mu}$ represent the perturbation of density, pressure, and four-vector velocity, and bar has been employed for the background quantities. Note that the anisotropic part of the above relation has not been considered, because its spatial part can be traceless. As a result, by choosing the anisotropic part of the perturbation of the energy-momentum tensor to be zero, the generality of the problem does not lose.
Now, assuming a non-relativistic and pressureless matter, i.e., $\omega_{\rm m} = p = 0$, as the dominant factor managing the dynamics of the Universe, non-zero components of Eq.~(\ref{energy-momentum-full}) are~\cite{Haghani}
\begin{equation} \label{pert-EM}
\bar{T}^0_0 = \rho,	\kern 2pc	\delta T^0_0 = - \rho \delta_{\rm m} ,	\kern 2pc \delta T^0_i = \rho \partial_i v,
\end{equation}
where $\delta_m=\delta \rho / \rho$ is the density contrast of matter and $v$ is the scalar mode of the velocity perturbation. Therefore, by inserting Eq.~(\ref{pert-EM}) into Eq.~(\ref{EM-full}), one can calculate the perturbed temporal and spatial components of the conserved energy-momentum tensor as
\begin{equation}
\xi = 3 \dot{\psi} - \dot{\delta}_{\rm m},
\label{nu0}
\end{equation}
and
\begin{equation}
\dot{\xi} + \mathcal{H} \xi - k^{2} \phi = 0,
\label{nui}
\end{equation}
where $\xi = \nabla_i \nabla^i v$ is the divergence of the velocity and $k$ is the wave number. Nevertheless, the definition of perturbed metric and energy-momentum tensor lead to the perturbed Einstein's field equations. In this regard, the (00) component of Eq.~(\ref{VarLE$GB$}) can be specified as
\begin{equation}
a^4 \rho \delta_{\rm m} + 4 \mathcal{P}(a) (k^2 \psi + 3 \mathcal{H}^2 \phi +3 \mathcal{H} \dot{\psi}) = 0.
\label{00component}
\end{equation}
Moreover, the spatial off-diagonal components of Eq.~(\ref{VarLE$GB$}) take the following form
\begin{equation}
\mathcal{P}(a) \phi + (\mathcal{Q}(a) - 4 \alpha \dot{\mathcal{H}}) \psi = 0.
\label{off-diagonal}
\end{equation}
and, the remaining components yield
\begin{equation}
\begin{array}{l}
2 \mathcal{H} \mathcal{P}(a) \dot{\phi} + \{ 4 \dot{\mathcal{H}} (2 \mathcal{P}(a) - a^2 \kappa^2) - k^2 \mathcal{P}(a) - 2 \mathcal{H}^2 \mathcal{Q}(a) \}\phi \\
+ 2 \mathcal{P}(a) \ddot{\psi} + 4 \mathcal{H} (a^2 \kappa^2 + 2 \alpha \dot{\mathcal{H}}) \dot{\psi} + k^2 ( 4 \alpha \dot{\mathcal{H}} - \mathcal{Q}(a) ) \psi = 0.
\end{array}
\end{equation}
where we have defined $\mathcal{P}(a)$ and $\mathcal{Q}(a)$ to be
\begin{equation}
\begin{array}{l}
\mathcal{P} (a)= 2 \alpha \mathcal{H}^2 + a^2 \kappa^2 , \\ \\
\mathcal{Q} (a)= 2 \alpha \mathcal{H}^2 - a^2 \kappa^2.
\label{PQ}
\end{array}
\end{equation}
It is worth noting that while we ignore the anisotropic stress tensor, one can expect that by taking the limit of $\alpha \rightarrow 0$, Eq.~(\ref{off-diagonal}) reduces to $\phi = \psi$, similar to the $\Lambda$CDM model.

\begin{figure}
\centering
\includegraphics[width=\columnwidth]{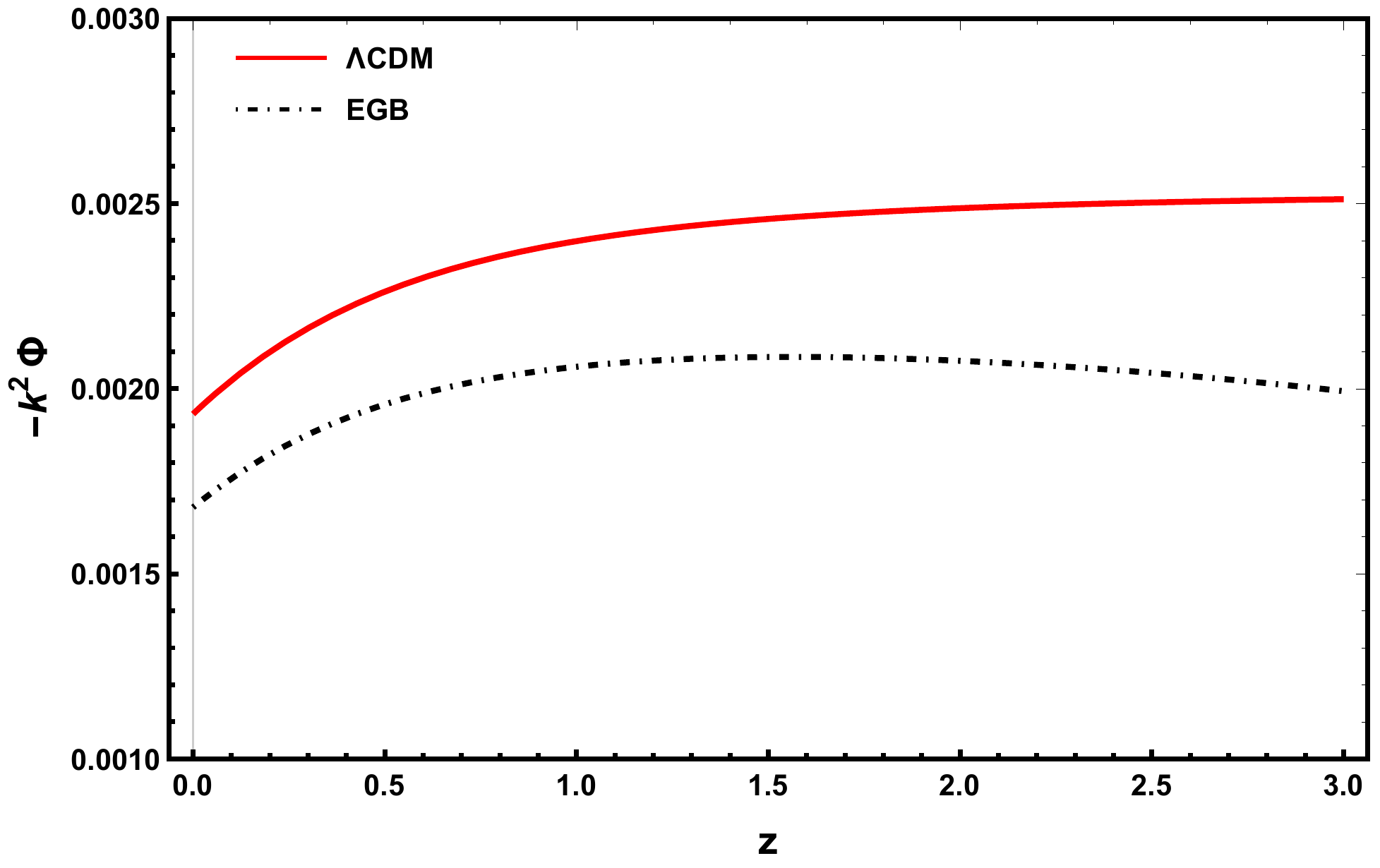}
\caption{The redshift-evolution of dimensionless gravitational potential for the 4D EGB and $\Lambda$CDM models. The dot-dashed (black) line indicates the calculations for the 4D EGB, while the solid (red) line shows the corresponding results for the $\Lambda$CDM model.}
\label{GP-fig}
\end{figure}

\begin{figure}
\centering
\includegraphics[width=\columnwidth]{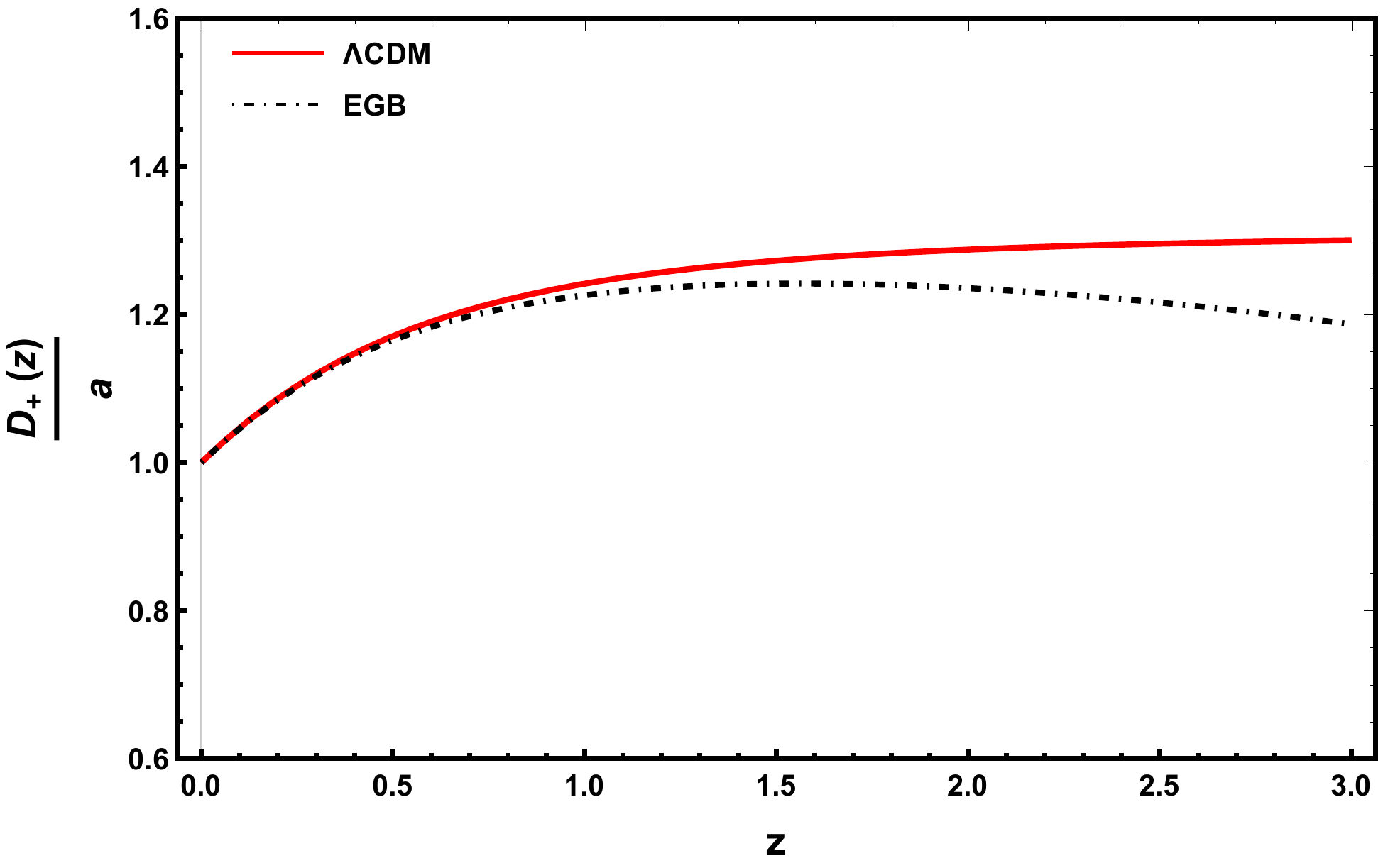}
\caption{The redshift-evolution of the linear growth factor normalized by the scale factor for the 4D EGB and $\Lambda$CDM models. The dot-dashed (black) line indicates this dependency for the 4D EGB model, while the solid (red) line shows the corresponding result for the $\Lambda$CDM model.}
\label{LGF-fig}
\end{figure}

It is believed that the Fourier modes can be divided into two parts~\cite{Uzan}. Super-Hubble modes, i.e., $k \ll H$, and sub-Hubble modes, i.e., $k \gg H$. Since the growing modes are of interest, we have considered the sub-Hubble limit. With this consideration, by substituting Eq.~(\ref{00component}) into Eq.~(\ref{off-diagonal}) the Poisson equation of the 4D EGB model takes the following form
\begin{equation}
4 \mathcal{P}(a)^2 k^2 \phi = a^4 (\mathcal{Q}(a) - 4 \alpha \dot{\mathcal{H}}) \rho \delta_m.
\label{Poisson-eq}
\end{equation}
Actually, by using this relation one can specify the evolution of the gravitational potential, $\phi$, for the 4D EGB model as a function of redshift. In Fig.~\ref{GP-fig}, we have depicted the redshift-evolution of the gravitational potential for the 4D EGB model, and have compared it with the corresponding result obtained from the $\Lambda$CDM model. As it is clear, the gravitational potential of the 4D EGB model is lower than the corresponding one obtained from the $\Lambda$CDM model for all values of redshifts. Since the ISW effect corresponds to the integrated change of the gravitational potentials, it is expected that this difference leads to a change in the related cosmological parameters to the 4D EGB model.
In addition, one can express the matter density contrast in terms of decaying modes and growing modes. On the large and small scales, we are only interested in growing modes, because the constructive and destructive interference of these modes lead to clusters or voids, respectively.
Additionally, all modes of interest have entered the Hubble horizon in the late-time Universe. In models with $\Omega_{\rm m0} \neq 1$, and with $\Omega_{\Lambda0}$, all modes experience the same growth factor in a way that the result leads to the same evolution for those modes~\cite{Dodelson}. Moreover, the linear growth factor can be defined as
\begin{equation}
D_{+} (z) = \frac{\delta_{\rm m} (z)}{\delta_{\rm m} (z=0)}.
\label{Growth-factor}
\end{equation}
By substituing Eq.~(\ref{off-diagonal}) into Eq.~(\ref{nu0}) and Eq.~(\ref{nui}), one can achieve a second-order differential equation for the linear growth factor as
\begin{equation}
\begin{array}{l}
D'' + \dfrac{E'}{E} D' \\ \\
\kern 1.4pc + \dfrac{3 \Omega_{m0} \{ 2 \beta E^2 (1+z)^2 + 4 \beta (1+z)^3 E E' -1 \}}{2 (1+z) \{2 \beta E^3 (1+z)^2 + E \}^2}D = 0,
\end{array}
\end{equation}
where prime denotes derivative with respect to the redshift $z$. To solve such an equation in general form, a numerical solution with proper initial conditions is required. For this purpose, we have considered the following initial condition
\begin{equation}
\frac{d D}{d \ln a} |_{z_*} = \gamma D |_{z_*},
\end{equation}
where $\gamma$ and $z_*$ are arbitrary constant and redshift. Note that the observational data must set $ \gamma$ and $\beta$. It should also be mentioned that we have used the best-fit value of parameters $ \gamma$ and $z_{*}$ obtained in Ref.~\cite{Haghani}, and have presented those along with other cosmological parameters in Table~\ref{tab:parameters}.
Fig.~\ref{LGF-fig} shows the normalized linear growth factor of the density contrast, i.e., $D_+ / a$, as a function of redshift for the 4D EGB and $\Lambda$CDM models. The result indicates that for the low redshifts, specifically for $z<1$, the shape of linear growth factor for the 4D EGB model tends to one obtained from the $\Lambda$CDM model in a way that at the present-time Universe, i.e., $z=0$, the growth factors of both models reach the same value. While the deviation of the 4D EGB growth factor from the corresponding result obtained for the $\Lambda$CDM model increases by increasing the redshift.

\begin{figure}
\centering
\includegraphics[width=\columnwidth]{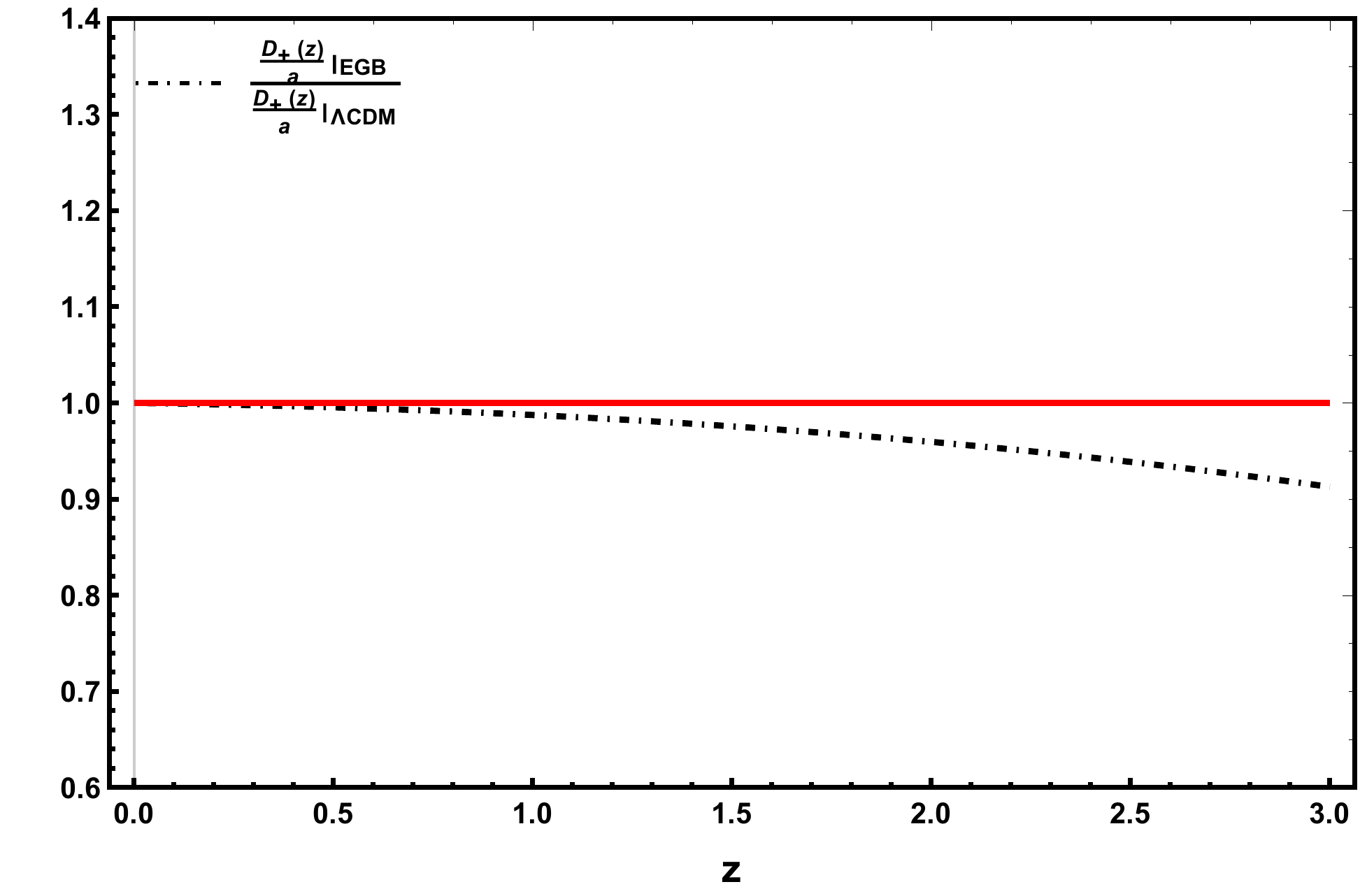}
\caption{Relative comparison between the redshift-evolution of the linear growth factor normalized by the scale factor for the 4D EGB model and the corresponding result obtained from the $\Lambda$CDM model with the amplitudes being normalized to unity at the present-time Universe (i.e., at $z = 0$). The solid (red) line has been depicted for comparison.}
\label{LGFN-fig}
\end{figure}

In Fig.~\ref{LGFN-fig}, we have indicated the ratio of $D_+/a$ in the 4D EGB model to the corresponding value obtained from the $\Lambda$CDM model as a function of redshift with the amplitudes being normalized to unity at the present-time Universe ($z = 0$) for better comparison between the results obtained from both models. Obviously, the value of this ratio is always lower than one, which informs that the normalized growth factor for the 4D EGB model is always lower than the one obtained for the $\Lambda$CDM model. 

It can be inferred from Figs.~(\ref{LGF-fig}) and (\ref{LGFN-fig}) that the amplitudes of the growth factor for the 4D EGB model is smaller than that of the $\Lambda$CDM one, which indicates that the 4D EGB model must have primordial perturbations smaller than the $\Lambda$CDM model to give the same quantity of structures at the present-time Universe.
\begin{table}
\centering
\caption{The best-fit parameters $\gamma$ and $z_{*}$, and the other cosmological parameters.}
\label{tab:parameters}
\begin{tabular}{c|c|c|c|c|c}
\hline
\hline
$\Omega_{c} h^2$ & $\Omega_{b} h^2$ & $H_0$ & $\Omega_{m0}$ & $\gamma$& $z_{*}$ \\
\hline
$0.112$ & $0.0226$ & $70$ & $0.275$ & $3.8652$ & $7.1$ \\
\hline
\hline
\end{tabular}
\end{table}
\section{The Integrated Sachs-Wolfe Effect}\label{sec:iv}
As the CMB photons travel from the last scattering surface to us, they move through gravitational potentials generated by the accumulated matters, e.g., the galaxy clusters. It is well known that the photons become blueshifted when they move into potential wells, while the photons will be redshifted as they move out of the gravitational potential wells. Accordingly, these shifts will be accumulated along the line of sight of the observer. In a Universe with no dark energy or curvature, within the linear regime of fluctuations, the gravitational potentials will not change with time if the expansion of the Universe is dominated by a fluid with a constant EoS. Therefore, during the matter-dominated era, the total shift of photons will cancel out. However, any deviation from the matter domination, i.e., the evolution of the EoS, causes the potentials to vary with time. As a result, an integrated change in the energy of traveling photons occurs as they pass through these evolving potentials. This phenomenon has been observed as anisotropy in the CMB temperature, called the ISW effect.
Moreover, the ISW contribution to the total CMB spectrum is only important on large scales (i.e., low CMB multipoles, $l$), where the possibility of extracting information is limited by cosmic variance. However, there are other ways to detect the ISW effect, which the most practical one is the measurement of the correlation between the CMB and the large scale structure \cite{Crittenden and Turok.}.
The ISW effect is important for two main reasons. First, it is a direct probe of dark energy properties, and second, it may be the source of some of the large-scale anomalies in the CMB temperature data \cite{1407.5623}. In the following sections, we mainly focus on the ISW effect and drive the auto-correlation and cross-correlation angular spectrum for the 4D EGB model, and compare those with the corresponding results obtained from the $\Lambda$CDM model.

\subsection{Gravitational Potential}
While a photon propagates toward a cosmological object from the last scattering surface, the gravitational potential evolve in two periods of time. First, in the early Universe, right after the recombination era due to the presence of the radiation, and second, in the late-time Universe due to the presence of dark energy~\cite{Dodelson2020}. As mentioned before, these evolving potentials lead to CMB temperature anisotropies. Consequently, the evolution of the gravitational potential plays a vital role for studying the ISW effect, namely
\begin{equation}
\bigg( \dfrac{\Delta T}{T_{\rm CMB}} \bigg)_{\rm ISW} = \dfrac{2}{c^2} \int^1_{a_{\rm dec}} \frac{\partial \phi}{\partial a} da,
\label{deltaT/T}
\end{equation}
where $T_{\rm CMB} = 2.725~\rm K$ is the CMB temperature and $a_{\rm dec}$ denotes the time at which the photon decoupling has occurred. It must be emphasized that photon decoupling has occurred at $t_{\rm dec}\simeq 378,000$ years after the big bang, or equivalently at redshift $z_{\rm dec}\simeq 1100$. We have also chosen the comoving coordinate systems, because in this system, cosmological fluids are at rest, and galaxies are spatially fixed \cite{Hobson}. In a flat Universe, the comoving distance between an object and us can be calculated as
\begin{equation}
\chi(a) = c \int_a^1 \frac{da}{a \mathcal{H}(a)},
\end{equation}
where $c$ is the velocity of light in vacuum. With this consideration, Eq.~(\ref{deltaT/T}) can be reformed as
\begin{equation}
\bigg( \dfrac{\Delta T}{T_{\rm CMB}} \bigg)_{\rm ISW} = - \dfrac{2}{c^3} \int_0^{\chi_{\rm H}} d \chi\,a^2 H(a) \frac{\partial \phi}{\partial a}.
\label{deltaT/T-n}
\end{equation}
On the other hand, the gravitational potential is determined by the following perturbed Poisson equation
\begin{equation}
- k^2 \phi = \dfrac{a^4 \Big( 4 \alpha \dot{\mathcal{H}} - \mathcal{Q}(a) \Big)}{4 \mathcal{P}^2(a)} \rho_{\rm m} \delta_{\rm m} (a),
\label{pert-Poisson}
\end{equation}
where $\rho_{\rm m}(a) = \Omega_{\rm m}(a) \rho_{\rm crit}$ is the mean background matter density, $\Omega_{\rm m}(a)$ is the matter density parameter and $\rho_{\rm crit}(a)$ is the critical density and can be specified through Eq.~(\ref{Friedmann-eq}). Furthermore, because the cosmological fluids evolve independent from each other, the matter density follows
\begin{equation}
\Omega_{\rm m} (a) = \dfrac{\Omega_{\rm m0} H_{\rm 0}^2}{a \mathcal{H}^2(a)}.
\end{equation}
Under these assumptions and by inserting Eq.~(\ref{Growth-factor}) into Eq.~(\ref{pert-Poisson}), one can achieve the following relation for the Poisson equation
\begin{equation}
\phi(k,a) = \dfrac{-1}{4 k^2} \mathcal{V}(a) \delta (k,a=1),
\label{pert-Poisson-n}
\end{equation}
where $\mathcal{V}(a)$ is a function of scale factor that is defined as
\begin{equation}
\begin{array}{l}
\mathcal{V}(a) \equiv \dfrac{a^4 (4 \alpha \dot{\mathcal{H}} - \mathcal{Q}(a))}{\mathcal{P}^2(a)} \times \\ \\
\kern 1.5pc \Bigg( \dfrac{2 \kappa^2 (3 \mathcal{H}^2 - a^2 \Lambda)}{a^2} + \dfrac{6 \alpha \mathcal{H}^4}{a^4} \Bigg) \Omega_{\rm m}(a) D_+(a).
\end{array}
\end{equation}
Therefore, substituting Eq.~(\ref{pert-Poisson-n}) into Eq.~(\ref{deltaT/T-n}) leads to the following form of the CMB temperature perturbation caused by the ISW effect
\begin{equation}
\bigg( \dfrac{\Delta T}{T_{\rm CMB}} \bigg)_{\rm ISW} = \dfrac{1}{2 c^3} \int_0^{\chi_H} d \chi ~ a \mathcal{H}(a) \dfrac{d \mathcal{V}}{d a} \dfrac{\delta(k,a=1)}{k^2}.
\end{equation}
In addition to the CMB temperature, we need information from the growth of structures. In order to achieve this purpose, one has to calculate the cross-correlation between the CMB and a foreground density field covering the entire extragalactic sky. On the other hand, the large-scale structure surveys, by counting the abundance of galaxy clusters as a function of redshift, are powerful probes of the growth of structure~\cite{0802.2522}. Moreover, the intrinsic angular galaxy fluctuations can be defined as~\cite{0801.0642}
\begin{equation}
\delta_{\rm g} = \int f(z) \delta_{\rm m}(z) dz ,
\end{equation}
where $f(z)=b(z)\,dN/dz$ is the redshift distribution function of the observed samples, $b(z)$ is the redshift-dependent bias that relates the baryonic matter to dark matter, and $dN/dz$ is the redshift distribution of the survey in the comoving distance. Furthermore, the comoving distance is related to the redshift via $d \chi =-(a c/ \mathcal{H})dz$. Under these assumptions, the intrinsic angular galaxy fluctuations takes the following form
\begin{equation}
\delta_{\rm g} = \int \dfrac{\mathcal{H}(a)}{ac}f(z)D_+(a)\delta_{\rm m}(k,a=1) d\chi .
\end{equation}
By utilizing these functions, we are able to calculate the ISW-auto spectrum~\cite{0803.2239, 1504.02416, 1004.3341}
\begin{equation}\label{ctt-spec}
C^{\rm ISW}_{\rm TT} (l) = \int_0^{\chi_{\rm H}}\dfrac{W^2_{\rm T} (\chi)}{\chi^2} \dfrac{H_0^4 P_{\delta \delta} (k=l/\chi)}{k^4} d \chi ,
\end{equation}
the the ISW-cross spectrum
\begin{equation}\label{ctg-spec}
C_{\rm Tg} (l) = \int_0^{\chi_{\rm H}}\dfrac{W_T (\chi) W_{\rm g} (\chi)}{\chi^2} \dfrac{H_0^2 P_{\rm \delta \delta} (k=l/\chi)}{k^2} d \chi ,
\end{equation}
and the observed galaxies-auto spectrum
\begin{equation}\label{cgg-spec}
C_{\rm gg} (l) = \int_0^{\chi_{\rm H}} \dfrac{W^2_{\rm g} (\chi)}{\chi^2} P_{\rm \delta \delta} (k=l/\chi) d \chi,
\end{equation}
where $P_{\rm \delta \delta}$ is the present matter power spectrum that can be spesified by using the definition of the transfer function~\cite{0803.2239}, $l$ is the multipole order, and $W_{\rm T}$ and $W_{\rm g}$ are the ISW and galaxy window functions, respectively. 

\begin{figure}
\centering
\includegraphics[width=\columnwidth]{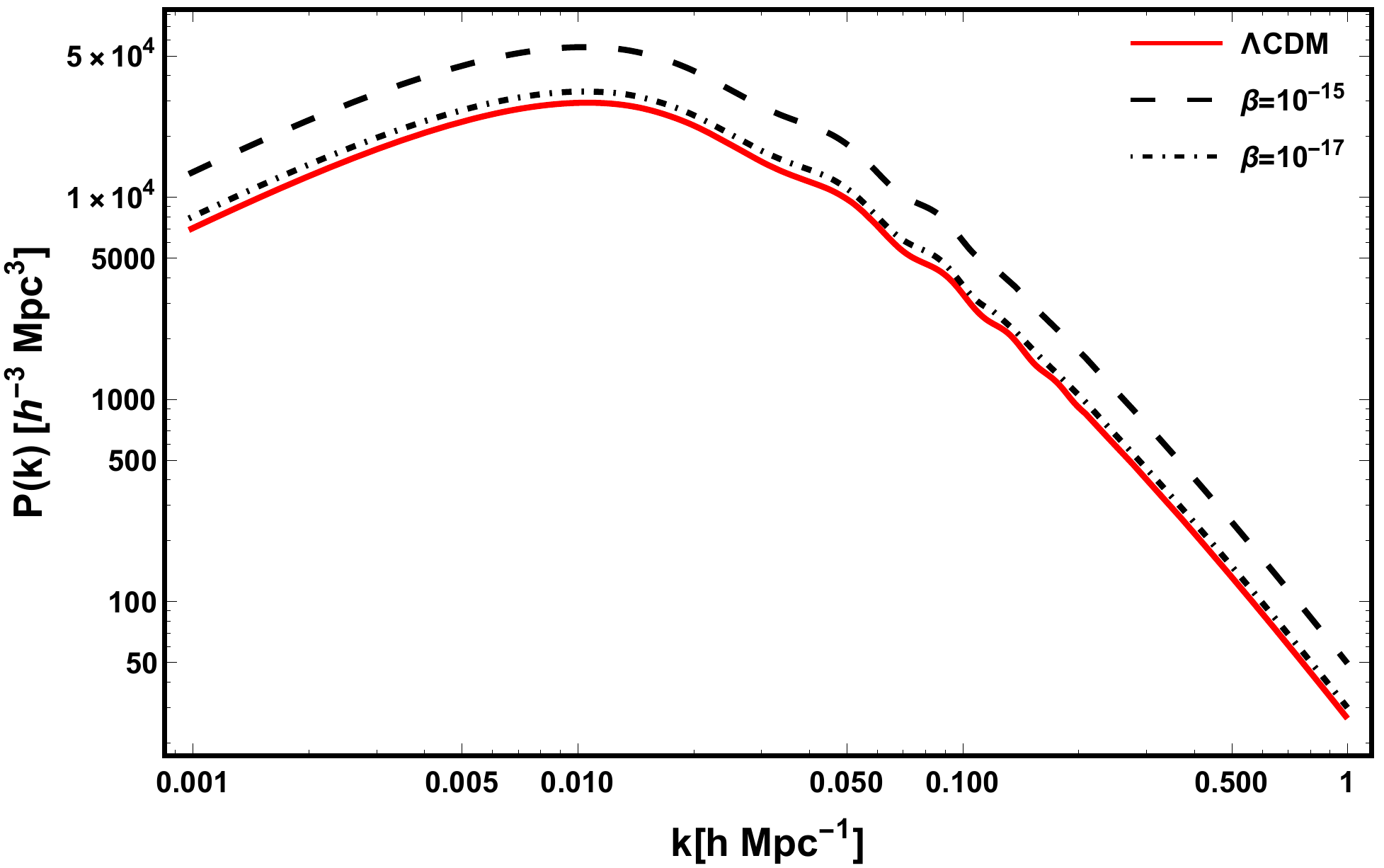}
\caption{The matter power spectrum as a function of wave number for the 4D EGB and $\Lambda$CDM models. The dashed (black) and the dot-dashed (black) lines indicate this dependency for the 4D EGB model while considering $\beta = 10^{-15}$ and $\beta = 10^{-17}$, respectively, whereas the solid (red) line shows the result for the $\Lambda$CDM model.}
\label{mps-fig}
\end{figure}

In Fig.~\ref{mps-fig}, we have shown the results obtained in Ref.~\cite{2103.12358} for the matter power spectrum as a function of wave number for the 4D EGB model while considering two values of $\beta$ and for the $\Lambda$CDM model. As can be seen from the figure, the matter power spectrum for the 4D EGB model is strengthened with respect to the one extracted from the $\Lambda$CDM model. Also, as expected, the matter power spectrum of the 4D EGB model tends to the corresponding result obtained for the $\Lambda$CDM model as the value of $\beta$ decreases. Nevertheless, for $\beta=10^{-15}$, 4D EGB has a larger power spectrum at all scales.

Other functions that have been introduced in Eqs.~(\ref{ctt-spec})-(\ref{cgg-spec}) are the window functions~\cite{0802.0983, 0803.2239}. The ISW window function in the case of a spatially flat Universe with non-clustering dark energy is
\begin{equation}
W_{\rm T} (\chi) = \dfrac{a}{2 c^3}\mathcal{H}(a) \dfrac{d \mathcal{V}}{d a},
\end{equation}
and the galaxy window function is
\begin{equation}
W_{\rm g} (\chi) = \dfrac{\mathcal{H}(a)}{ac} f(z) D_{+} (a).
\label{Wg-eq}
\end{equation}
The ISW effect depends on the redshift distribution of astronomical surveys that requires a lot of precise works, see, e.g., Ref.~\cite{0801.0642}. In this work, we have employed the latest results of some renowned surveys introduced in Refs.~\cite{1004.3341, 0803.2239, 1502.01595}.
Note that in the above formulas, all scale factors and redshifts must be converted into the comoving distance where the current size of the observable Universe is given approximately by the present-day Hubble distance $d_{H,0} = c H_0^{-1}$.
\begin{figure}
\centering
\includegraphics[width=\columnwidth]{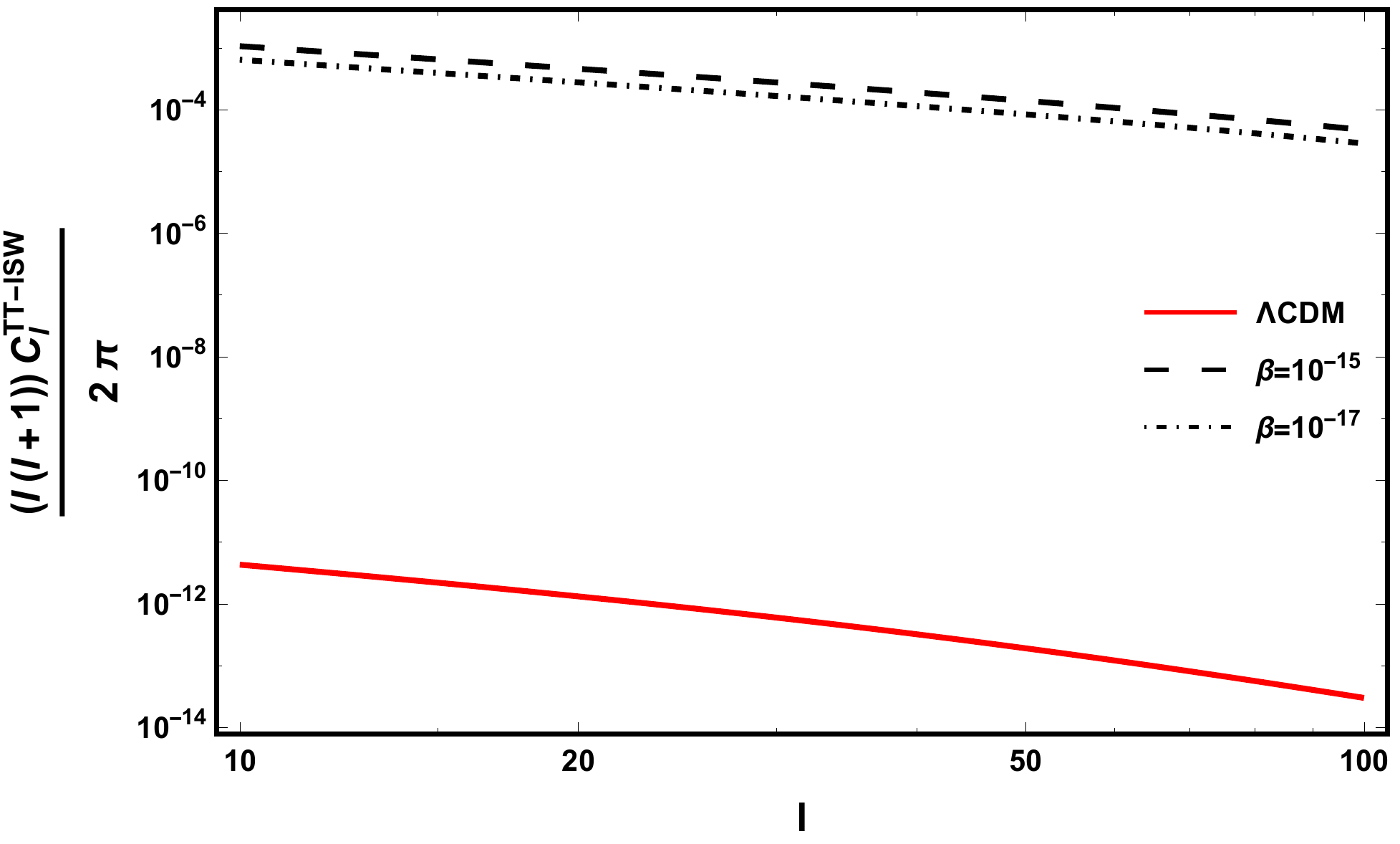}
\caption{The ISW-auto power spectrum as a function of multipole order $l$ for the 4D EGB and $\Lambda$CDM models. The dashed (black) and the dot-dashed (black) lines indicate this dependency for the 4D EGB model while considering $\beta = 10^{-15}$ and $\beta = 10^{-17}$, respectively, whereas the solid (red) line shows the result for the $\Lambda$CDM model.}
\label{CTT-fig}
\end{figure}
\subsection{Large-Scale Structure Surveys}
In order to measure the ISW effect, one has to find the cross-correlation between CMB maps and survey catalogs. In this section, we briefly review three galaxy catalogs and their analytical redshift distribution functions used in our calculation.

In Fig.~\ref{CTT-fig}, we have shown the ISW-auto power spectrum for the 4D EGB model, and have compared it with the result of the $\Lambda$CDM model. Obviously, the amplitude of the ISW-auto power spectrum for the 4D EGB model is higher than the one obtained from the $\Lambda$CDM model. Also, there is a major difference between the amplitude of the ISW-auto power spectrum of the 4D EGB model and the one extracted from the $\Lambda$CDM model. Since the total CMB temperature-auto power spectrum in the $\Lambda$CDM model is consistent with the observational results from the Planck satellite, a large increase in the ISW-auto power spectrum in the 4D EGB model may lead to the trouble. Although it is difficult to directly detect the amplitude of the ISW-auto power spectrum apart from the total CMB temperature anisotropies, this result may indicate the need for stronger observational constraints on the coupling parameter of the 4D EGB model.

\subsubsection{DUNE Survey}
The Dark Universe Explorer (DUNE) is a wide-field space imager that consists of a 1.2 m telescope and is designed to detect both visible and three near-infrared bands. It is optimized for weak gravitational lensing and the ISW effect as a complementary cosmological probe \cite{0802.0983}. For this survey, the redshift distribution function is defined as
\begin{equation}
f_{\rm DUNE}(z)=b_{\rm eff} \left[ \dfrac{z_*}{\alpha_*} \Gamma(\dfrac{3}{\alpha_*}) \right]^{-1} \bigg( \dfrac{z}{z_*} \bigg)^{2} \exp \left[-\left( \dfrac{z}{z_*} \right)^{\alpha_*}\right].
\end{equation}
where $\Gamma(x)$ is the Gamma funcion, and $b_{\rm eff}$, $z_{*}$ and $\alpha_{*}$ are free parameters to be specified.

\begin{figure}
\centering
\includegraphics[width=\columnwidth]{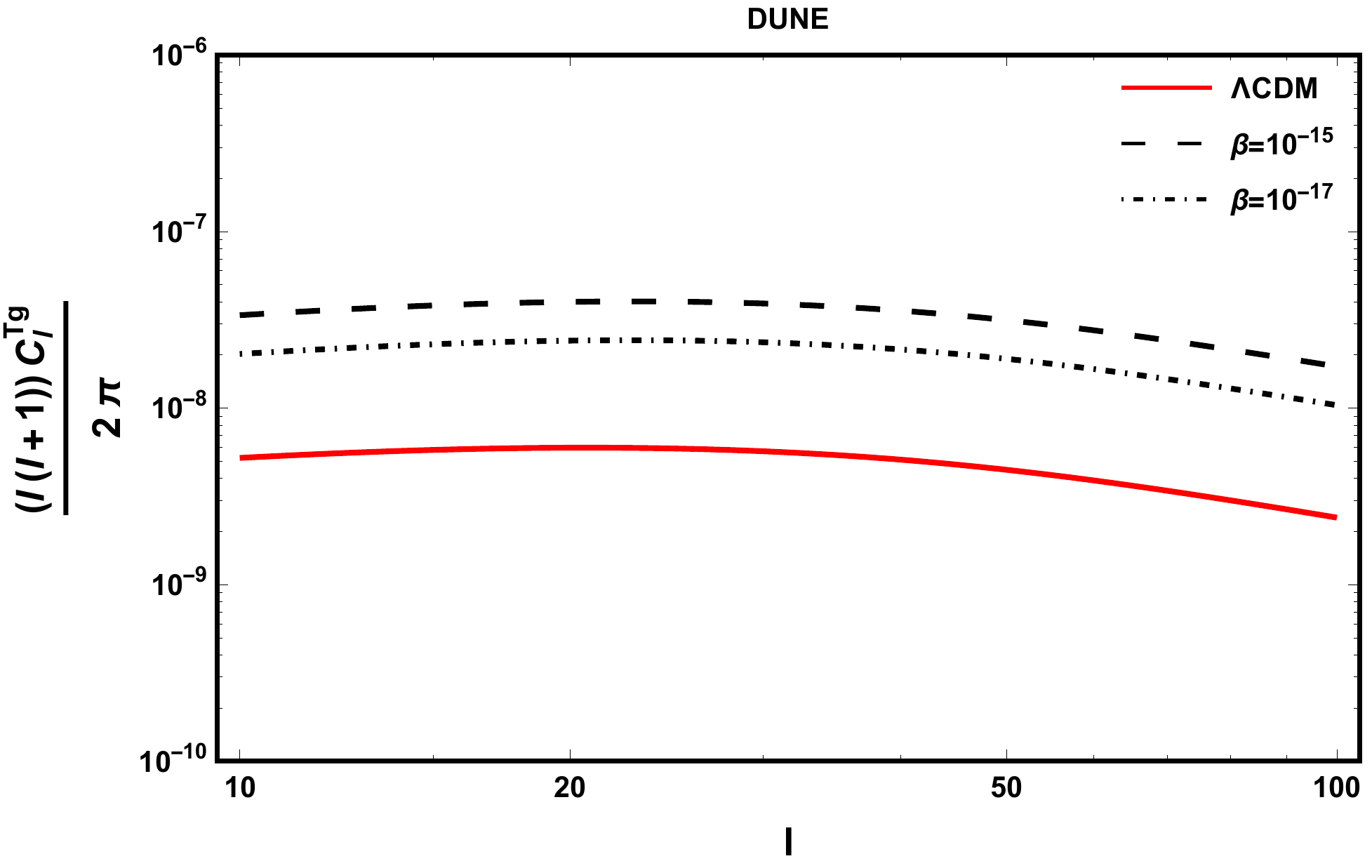}
\caption{The ISW-cross power spectrum as a function of multipole order $l$ for the 4D EGB and $\Lambda$CDM models. The dashed (black) and the dot-dashed (black) lines indicate this dependency for the 4D EGB model while considering $\beta = 10^{-15}$ and $\beta = 10^{-17}$, respectively, whereas the solid (red) line shows the result for the $\Lambda$CDM model. The results have been depicted for the DUNE sample.}
\label{CTg-DUNE-fig}
\end{figure}

In Fig.~\ref{CTg-DUNE-fig}, for the DUNE survey, we have shown the ISW-cross power spectrum as a function of multipole order, $l$, while considering the 4D EGB model, and have compared it with the corresponding result of the $\Lambda$CDM model. As can be seen from the figure, the amplitude of the ISW-cross power spectrum for the 4D EGB model is higher than the one obtained from the $\Lambda$CDM model. The results indicate that the 4D EGB model can amplify the ISW-cross power spectrum, which can be considered as a relative advantage of the 4D EGB model. Moreover, it is clear that the deviation from the $\Lambda$CDM model is directly proportional to the value of $\beta$.

\subsubsection{NVSS Survey}
In addition to visible light, radio astronomy is relatively deep and wide in a way that it provides an excellent opportunity to study large-scale structures and especially the ISW effect. The National Radio Astronomy Observatory (NRAO) Very Large Array (VLA) Sky Survey (NVSS) is a $1.4~\rm GHz$ continuum survey that is covering about $82\%$ of the sky, the entire sky north of $-40~\rm deg$ declination~\cite{NVSS, 1203.3277}. Finding a redshift distribution function for NVSS has difficulties. The common method is by cross-correlation against the other samples whose redshift distributions are known. Albeit this method has some flaws, for instance, there are limited data for $z>2.6$ due to the limitation of the range of the other samples \cite{1407.5623, NVSS}. For this survey, the redshift distribution function is defined as
\begin{equation}
f_{\rm NVSS} (z) = b_{\rm eff}\, \dfrac{\alpha_*^{\alpha_* + 1}}{\Gamma (\alpha_*)} \frac{z^{\alpha_*}}{z_*^{\alpha_* +1}}\exp \left( -\frac{\alpha_* z}{z_*} \right).
\end{equation}

\begin{figure}
\centering
\includegraphics[width=\columnwidth]{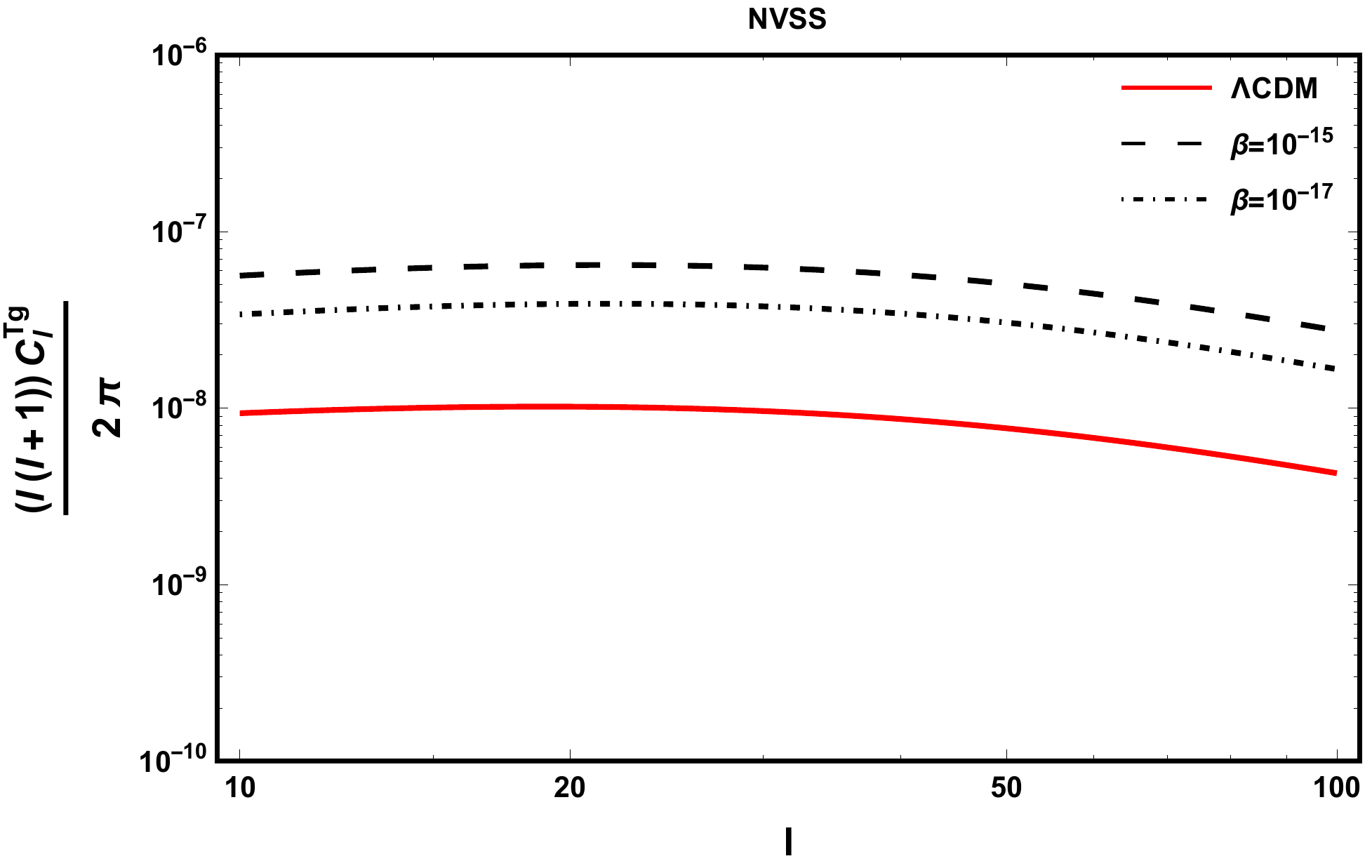}
\caption{The ISW-cross power spectrum as a function of multipole order $l$ for the 4D EGB and $\Lambda$CDM models. The dashed (black) and the dot-dashed (black) lines indicate this dependency for the 4D EGB model while considering $\beta = 10^{-15}$ and $\beta = 10^{-17}$, respectively, whereas the solid (red) line shows the result for the $\Lambda$CDM model. The results have been depicted for the NVSS sample.}
\label{CTg-NVSS-fig}
\end{figure}

In Fig.~\ref{CTg-NVSS-fig}, for the NVSS survey, we have shown the ISW-cross power spectrum as a function of multipole order, $l$, for the 4D EGB model, and have compared it with the corresponding result of the $\Lambda$CDM model. Clearly, the results exhibit that the 4D EGB model can increase the amplitude of the ISW-cross power spectrum in a way that its deviation from the results of $\Lambda$CDM model changes directly with the value of the dimansionless coupling parameter $\beta$.

\begin{table}
\centering
\caption{The best-fit parameters of the redshift distribution of three different surveies, i.e., DUNE, NVSS and SDSS surveys.}
\label{tab-best-fit}
\begin{tabular}{c|c|c|c|c}
\hline
\hline
Survey~~ & ~~$b_{\rm eff}$~~ & ~~$z_*$ & ~~$\alpha_*$~~ & ~~$m$~~ \\
\hline
DUNE~~ & ~~$1.00$~~ & ~~$0.640$~~ & ~~$1.500$~~ & ~~$ - $~~ \\
\hline
NVSS~~ & ~~$1.98$~~ & ~~$0.790$~~ & ~~$1.180$~~ & ~~$ - $~~ \\
\hline
SDSS~~ & ~~$1.00$~~ & ~~$0.113$~~ & ~~$1.197$~~ & ~~$3.457$~~ \\
\hline
\hline
\end{tabular}
\end{table}

\subsubsection{SDSS Survey}
The Sloan Digital Sky Survey (SDSS) is an imaging and spectroscopic survey that collect galaxies, quasars, and stars. There are different subsamples of SDSS. In this work, we have used the redshift distribution of the main photometric SDSS galaxy sample (SDSS-MphG)
that has been introduced in Ref.~\cite{1004.3341}. For this survey, the redshift distribution function is
\begin{equation}
f_{\rm SDSS} (z) = b_{\rm eff}\, \dfrac{\alpha_*}{\Gamma (\frac{m+1}{\alpha_*})} \frac{z^m}{z_*^{m+1}} \exp \left[ - \left(\dfrac{z}{z_*}\right)^{\alpha_*} \right].
\end{equation}
where $m$ is a free parameter to be determined. We have also presented the best-fit values of the redshift distribution of these three surveys in Table~\ref{tab-best-fit}.

\begin{figure}
\centering
\includegraphics[width=\columnwidth]{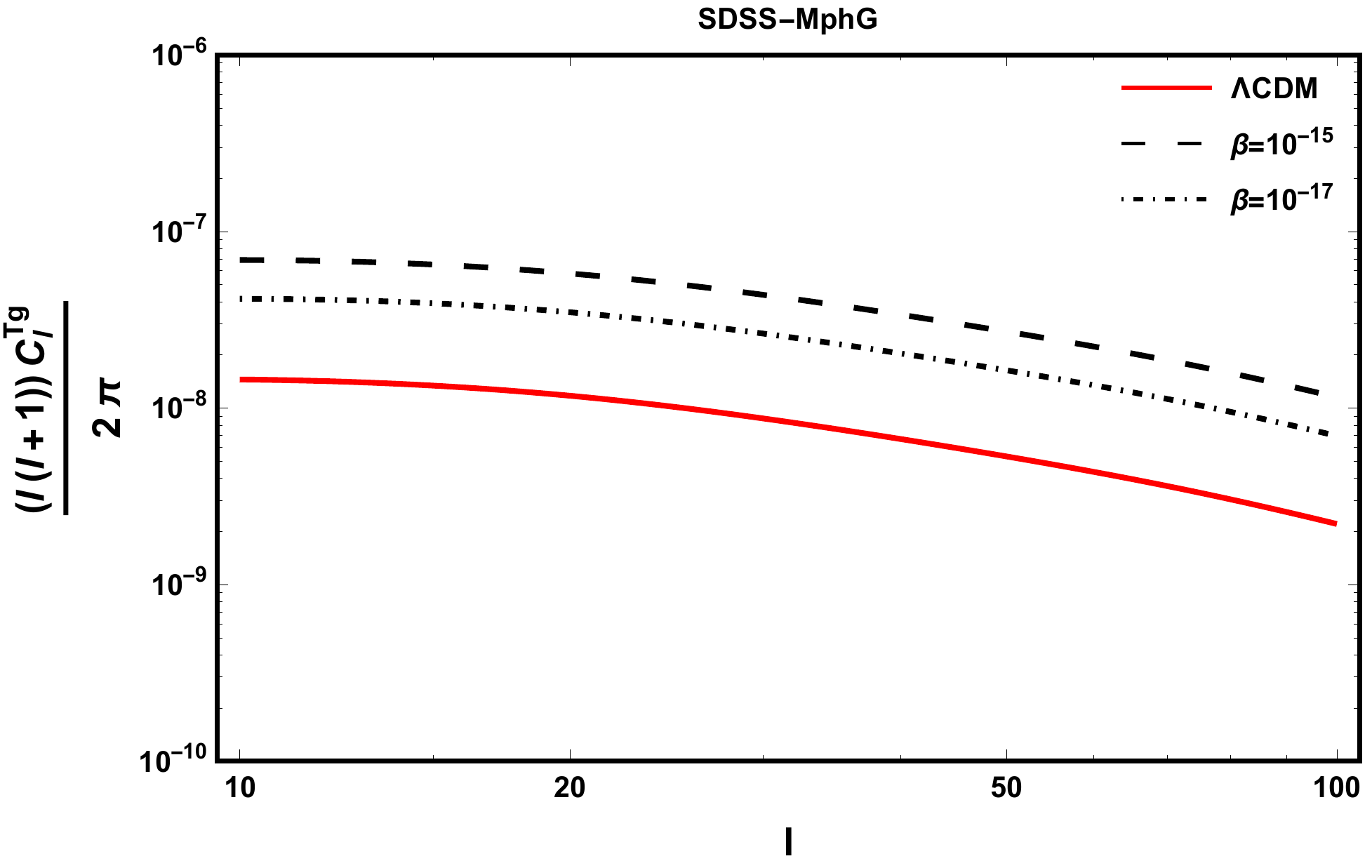}
\caption{The ISW-cross power spectrum as a function of multipole order $l$ for the 4D EGB and $\Lambda$CDM models. The dashed (black) and the dot-dashed (black) lines indicate this dependency for the 4D EGB model while considering $\beta = 10^{-15}$ and $\beta = 10^{-17}$, respectively, whereas the solid (red) line shows the result for the $\Lambda$CDM model. The results have been depicted for the SDSS-MphG sample.}
\label{CTg-SDSS-fig}
\end{figure}

In Fig.~\ref{CTg-SDSS-fig}, for the SDSS survey, we have indicated the ISW-cross power spectrum as a function of multipole order for the 4D EGB model, and have compared it with the corresponding result of the $\Lambda$CDM model. As can be seen, the results are in good agreement with corresponding ones obtained from the DUNE and NVSS surveys.

The obtained findings from these three surveys, confirm that the amplitude of the ISW-cross power spectrum for the 4D EGB model is higher than the one obtained for the $\Lambda$CDM model. Moreover, the results indicate that the ISW-cross power spectrum for the 4D EGB model will tend to the corresponding value extracted from the $\Lambda$CDM model if the dimansionless coupling parameter takes the smaller values.

\begin{figure}
\centering
\includegraphics[width=\columnwidth]{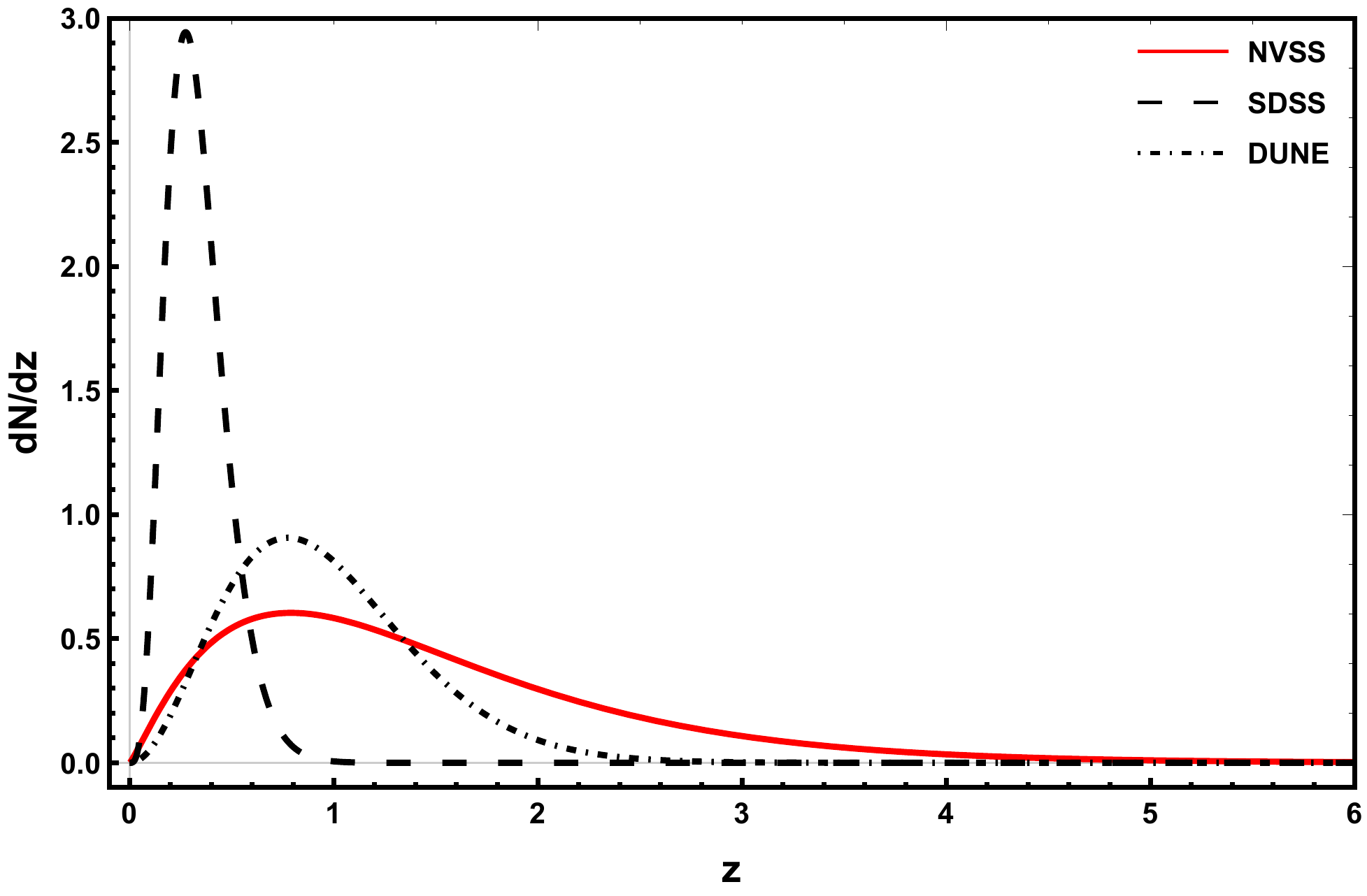}
\caption{The redshift distribution for different surveys. The solid (red) line, the dot-dashed (black) line, and the dashed (black) line indicate the corresponding results for the NVSS, DUNE, and SDSS surveys.}
\label{RD-fig}
\end{figure}

In Fig.~\ref{RD-fig}, we have also depicted the normalized redshift distribution of the mentioned surveys as a function of redshift. As can be seen from the figure, the NVSS survey offers the widest redshift coverage with respect to the other surveys.

\section{Conclusions}\label{sec:v}
The EGB gravity was initially proposed as an alternative for dark energy to explain the expansion of the Universe. But contrary to expectations, the theory does not participate in the gravitational dynamics in 4D, because the GB invariant is a total derivative. In fact, the variation of the GB action with respect to the metric is identically zero in 4D due to the presence of an overall factor $(D-4)$. In this regard, Glavan and Lin proposed the novel 4D EGB gravity by rescaling the coupling constant $\alpha \rightarrow \alpha / (D-4)$ to yield a non-trivial contribution to the gravitational dynamics.

In this work, we have calculated the ISW effect in the 4D EGB gravity. For this purpose, we have initially discussed the theoretical framework of the 4D EGB gravity and different cosmological aspects of this theory. First, we have indicated the behavior of the gravitational potentials in the 4D EGB gravity. The results illustrate that the gravitational potential of the 4D EGB model is lower than the corresponding value obtained from the $\Lambda$CDM model for all values of redshift. It has also been confirmed that the deviation of the gravitational potential of the 4D EGB model from the one obtained from the $\Lambda$CDM model changes directly with the redshift.

Moreover, we have calculated the linear growth factor of the 4D EGB model, and have compared it with the corresponding result obtained for the $\Lambda$CDM model. The results demonstrate that for the low redshifts, specifically for $z<1$, the shape of the linear growth factor for the 4D EGB model tends to the one obtained from the $\Lambda$CDM model in a way that at the present-time Universe, i.e., $z=0$, the growth factors of both models reach the same value. Also, the deviation of the 4D EGB growth factor from the corresponding result obtained for the $\Lambda$CDM model changes directly with the redshift. We have also shown the ratio of $D_+/a$ for the 4D EGB model to the one obtained from the $\Lambda$CDM model as a function of redshift for better comparison. The results represent that the values of the growth factor for the 4D EGB model should be smaller than those of the $\Lambda$CDM model as the growth factor is normalized at the present-time Universe, Hence, it indicates that the 4D EGB model must have primordial perturbations smaller than those in the $\Lambda$CDM model to give the same number of structures at the present-time Universe.

Also, we have shown the matter power spectrum with respect to the wave number for the 4D EGB model while considering two values of $\beta$ and for the $\Lambda$CDM model that have been presented in Ref.~\cite{2103.12358}. The results indicate that the matter power spectrum for the 4D EGB model is strengthened with respect to the one extracted from the $\Lambda$CDM model. Also, it has been confirmed that the matter power spectrum of the 4D EGB model tends to the one obtained for the $\Lambda$CDM model as the value of $\beta$ tends to zero.

Furthermore, we have calculated the ISW-auto power spectrum for the 4D EGB model, and have compared it with the corresponding result of the $\Lambda$CDM model. The results indicate that the amplitude of the ISW-auto power spectrum for the 4D EGB model should be higher than the one obtained from the $\Lambda$CDM model. Moreover, the results display a major difference between the amplitude of the ISW-auto power spectrum for the 4D EGB model and the one extracted from the $\Lambda$CDM model. Since the total CMB temperature-auto power spectrum in the $\Lambda$CDM model is consistent with the observational results from the Planck satellite, a large increase in the ISW-auto power spectrum in the 4D EGB model may not be suitable. This result may mean the need to place stronger observational constraints on the coupling parameter of the 4D EGB model.

Additionally, by employing three different surveys, we have calculated the ISW-cross power spectrum as a function of multipole order while considering the 4D EGB model, and have compared those results with the corresponding findings from the $\Lambda$CDM model. The results exhibit that the amplitude of the ISW-cross power spectrum for the 4D EGB model must be higher than the one obtained from the $\Lambda$CDM model. This means that the 4D EGB model can amplify the ISW-cross power spectrum, which can be considered as a relative advantage of the 4D EGB model. Moreover, it has been confirmed that the deviation of the 4D EGB ISW-cross power spectrum from the one obtained for the $\Lambda$CDM model is directly proportional to the value of the dimensionless coupling parameter $\beta$.

Finally, we have demonstrated the normalized redshift distribution of the DUNE, NVSS, and SDSS surveys as a function of redshift. The results indicate that the NVSS survey yields the widest redshift coverage with respect to the other surveys.

\section*{ACKNOWLEDGEMENT}
M.G. and A.B. warmly thank Abdolali Banihashemi for many useful discussions. M.G. gratefully acknowledges Farbod Hassani for invaluable comments on the numerical calculations. The authors would also like to appreciate Deng Wang who provided us with the 4D EGB matter power spectrum data.

\end{document}